\documentclass[reprint,twocolumn,superscriptaddress,showkeys,
nofootinbib,notitlepage,amsmath,amssymb,floatfix, aps, pra]{revtex4-2} % aps and pra were added temporarily to make the warning go away
\usepackage{latexsym}
\usepackage{amssymb}
\usepackage{lmodern}
\usepackage{url}
\usepackage{natbib}
\usepackage[dvipsnames]{xcolor}
\usepackage{multirow}
\usepackage{tikz,pgfplots,circuitikz}
\usepackage{braket}
\usepackage{esvect}
\usepackage{mathtools}
\usepackage[normalem]{ulem}
\usepackage{bm}
\usepackage{color} %red, green, blue, yellow, cyan, magenta, black, white
\newcommand{\rmd}{\mathrm{d}}

\newcommand{\Z}{\mathbb{Z}}

\newcommand{\mcD}{\mathcal{D}}

\renewcommand{\vv}[1]{\bm{\mathrm{#1}}}

\newcommand{\grad}{\bm{\mathrm{\nabla}}}

\newcommand{\Id}{1}

\pgfplotsset{compat=1.18, trig format plots = rad}
\usepgfplotslibrary{colorbrewer}

\def\de{\delta{\vphantom{\bar\delta}}}
\def\bde{{\bar\delta}}

\begin{document}
\title{The $\chi_{c1}(3872)$ Multiplet in the Diabatic Dynamical Diquark Model: \\ Meson Threshold Spin Dependence}
\author{Richard F. Lebed}
\email{richard.lebed@asu.edu}
\author{Ethan H. Pham}
\email{ehpham@asu.edu}
\affiliation{Department of Physics, Arizona State University, Tempe,
AZ 85287, USA}
\date{November, 2026}

\begin{abstract}
    We apply a minimal form of the diabatic dynamical diquark model, which uses only dimeson thresholds augmented by constraints obtained from heavy-quark spin symmetry, to the hidden-charm tetraquark spectrum.  We find that this model is insufficient to simultaneously reproduce the experimentally established states \(\chi_{c1}(3872)\), \(T_{c\bar c1}(3900)\), and \(T_{c\bar c1}(4020)\).  We therefore extend the model to its complete form, by restoring the spin- and isospin-dependent diquark-antidiquark interactions from the original adiabatic dynamical diquark model into the full diabatic potential.  Aligning the resulting framework to these three states allows the remaining members of the \(\Sigma_g^+(1S)\) multiplet to be predicted through coupled-channel scattering simulations.  The resulting resonance or bound state locations lie within approximately \(10~\mathrm{MeV}\) of the corresponding pure adiabatic mass eigenvalue predictions in all cases, indicating that the adiabatic fine-structure pattern survives the inclusion of explicit dimeson channels.  We also discuss possible experimental counterparts for the unconfirmed multiplet members.
\end{abstract}

\maketitle

\section{Introduction}\label{sec:Intro}

The discovery of numerous heavy-quark exotics over the past two decades has established a rich spectrum of states that cannot be accommodated within the conventional \(q\bar{q}\) meson and \(qqq\) baryon classifications.  More than 70 heavy-quark exotic candidates have been reported~\cite{ParticleDataGroup:2026mpi} at the time of writing, and yet we still do not understand the basic underlying structure of these states.  They nonetheless remain firmly within the realm of quantum chromodynamics (QCD), and their properties follow from the Standard Model.  In practice, however, the strongly coupled nature of QCD at hadronic energy scales makes direct calculations difficult, while first-principles lattice-QCD treatments of multihadron systems remain computationally demanding, even more so than for conventional hadrons.  Consequently, a variety of structural descriptions have been developed, including hadronic molecules, hybrids, hadroquarkonium, and compact multiquark configurations such as diquark-antidiquark states.

A recurring feature of the exotic spectrum is the proximity of many of these states to di-hadron thresholds.  For example, the state \(\chi_{c1}(3872)\) [previously \(X(3872)\)] lies exceptionally close to the \(D^{0}\bar{D}^{\ast 0}\) threshold.  Such cases indicate that coupling to nearby hadronic channels can play an important role in the observed spectrum, thus motivating descriptions capable of including effects due to both compact substructure and to nearby di-hadron channels.

Among the various approaches to understanding these exotics, the present work employs the dynamical diquark model~\cite{Lebed:2017min}.  This framework describes exotic hadrons in terms of colored quasiparticles, namely, diquarks \(\de=(q_1, q_2)\) and antidiquarks \(\bde=(\bar q_1, \bar q_2)\), with $\de\bde$ interaction dynamics organized by Born-Oppenheimer (BO) potentials, and has been developed for both tetraquark and pentaquark configurations (the latter also introducing colored triquarks $\theta$~\cite{Lebed:2015tna}).  Of particular interest in this work is the resulting organization of exotics into multiplets, which provides a systematic framework for relating experimentally observed states to additional unconfirmed spin and isospin partners, rather than treating each candidate independently.  The present work is restricted to hidden-charm tetraquarks, with the detailed $\de\bde$ construction reviewed in Sec.~\ref{sec:DynamicalDiquarkModel}.  The multiplet structure also leads to a comparatively rich predicted spectrum.  Even the BO ground-state tetraquark multiplet \(\Sigma_{g}^{+}(1S)\) contains six isosinglet and six isotriplet states, with still larger numbers appearing in the \(1P\) multiplet~\cite{Giron:2020fvd}, and additional observed states assigned to the \(2S\) excitations~\cite{Giron:2019bcs}.  This proliferation of predicted states has long presented a significant phenomenological obstacle for diquark-based descriptions, since many of the expected partners have not yet been clearly identified experimentally.  Their absence from the established spectrum does not imply that such states are spurious, however, as individual members may couple weakly to accessible channels, or simply be difficult to distinguish from hadronic backgrounds.  Identifying the remaining members of the lowest multiplet therefore provides an important test of the model.

The proximity of many exotic candidates to di-hadron thresholds motivates an extension of the original adiabatic picture in which the compact $\de\bde$ configurations may mix explicitly with nearby dihadron channels.  This synthesis is provided by the \textit{diabatic} dynamical diquark formalism~\cite{Lebed:2022vks,Lebed:2023kbm,Lebed:2024rsi,Lebed:2025xbz}, which permits the resulting states to arise through coupled-channel scattering.  The diabatic formalism may be further resolved by imposing heavy-quark spin symmetry (HQSS) upon heavy dihadron thresholds~\cite{Lebed:2025eqm}.  For a given \(J^{PC}\), the \(\delta\bar{\delta}\) state couples only to meson pairs of open heavy flavor whose partial-wave channels are capable of producing the same quantum numbers (\(D^{(\ast)}\bar{D}^{(\ast)}\) in the hidden-charm case).  The resulting potential matrices therefore encode both the relevant HQSS-constrained dimeson thresholds and their mixing with the \(\delta\bar{\delta}\) channel.

The present study focuses upon the hidden-charm \(c\bar{c}q\bar{q}'\) \(\Sigma_{g}^{+}(1S)\) multiplet containing \(\chi_{c1}(3872)\), \(T_{c\bar{c}1}(3900)\), and \(T_{c\bar{c}1}(4020)\).  Together, these states probe different spin and isospin sectors, as well as distinct open-charm thresholds, and thus provide a crucial set of experimental inputs for the model.   Their measured masses according to the Particle Data Group (PDG)~\cite{ParticleDataGroup:2026mpi} are:
\begin{align}\label{eq:PDGmasses}
    m_{\chi_{cq}(3872)}&=3871.7 \pm 0.17 \; \mathrm{MeV} , \notag\\
    m_{T_{c\bar{c}1}(3900)}&=3886.7\pm 2.3 \ \ \mathrm{MeV} , \notag\\
    m_{T_{c\bar{c}1}(4020)}&=4024.1\pm 1.9 \ \ \mathrm{MeV} .
\end{align}
Previous adiabatic and diabatic implementations emphasize complementary aspects of the problem: The former provides the internal spin-dependent fine structure of the multiplet in terms of direct $\de\bde$ couplings, while the latter incorporates the open dimeson channels relevant to its decay and scattering properties.  The present work combines these ingredients in order to investigate the resonance spectrum of the complete \(c\bar{c}q\bar{q}'\) \(\Sigma_{g}^{+}(1S)\) multiplet.

A complementary application of the diabatic BO approach has recently been applied to \(\chi_{c1}(3872)\) and its HQSS partners in Ref.~\cite{Alasiri:2026vfz}, and extended to the pentaquark sector in Ref.~\cite{Alasiri:2026ckx}.  While sharing the coupled-channel BO perspective employed here, that work uses a distinct construction of the states ({\it i.e.}, light quarks are incorporated directly into the diabatic potentials) and a different numerical treatment of the resulting scattering problem.  The present analysis instead employs the diabatic dynamical diquark formulation described above, in which an explicit compact \(\de\bde\) channel is coupled to the relevant open-charm dimeson thresholds.

The remainder of this work is organized as follows: Section~\ref{sec:DynamicalDiquarkModel} introduces the dynamical diquark model and its diabatic extension, as well as the relevant heavy-quark spin interactions, and examines whether the diabatic model with HQSS imposed upon its $D^{(*)} \bar D^{(*)}$ coupling (where HQSS breaking is taken to occur only through the $D$-$D^*$ mass difference) can reproduce the observed exotic spectrum.  Section~\ref{sec:AdiabaticCorrection} incorporates adiabatic fine-structure corrections for the $\de\bde$ component into the diabatic framework, compares the predicted \(\Sigma_{g}^{+}(1S)\) spectrum with that of the pure adiabatic model, and Sec.~\ref{sec:Disc} discusses possible observed counterparts for the remaining multiplet members.  Finally, Sec.~\ref{sec:Conclusion} summarizes the principal results and discusses their implications.

\section{The Dynamical Diquark Model}\label{sec:DynamicalDiquarkModel}

The dynamical diquark model~\cite{Lebed:2017min} introduces compact diquark $\de\equiv(qQ)_{\bar{\vv{3}}}$ and antidiquark $\bde\equiv(\bar{q}\bar{Q})_{\vv{3}}$ quasiparticles assumed to transform as color triplets (the unique attractive color combinations for such configurations at short distance~\cite{Brodsky:2014xia}).  $Q$ in the original model represents a heavy quark flavor ($c,b$), although the model has now been successfully extended into the hidden-strange sector~\cite{Jafarzade:2025qvx,Jafarzade:2025txm,Jafarzade:2026ogf}.  In some fraction of high-energy processes, $\de$ and $\bde$ form promptly with a high relative momentum; the $\de\bde$ pair then separates quickly, transferring its kinetic energy into the potential energy of the color flux tube between them.  This structure, prior to its hadronization through the wave-function overlap of the quarks in $\de$ with the antiquarks in $\bde$ to form either $(Q\bar q)(q\bar Q)$ or $(Q\bar Q)(q\bar q)$ meson pairs, constitutes the exotic state.

In this work, we consider \(c\bar{c}q\bar{q}'\) tetraquarks with \(q,q'\in\left\{u,d\right\}\), and assume that they are in the lowest Born-Oppenheimer (BO) potential \(\Sigma_{g}^{+}\left(1S\right)\) ({\it i.e.}, no excitation of the color flux tube), and lack internal orbital angular momentum.  Specific combinations of $\de$ and $\bde$ spins (denoted by $s^{\vphantom{-}}_\de$ and $s_\bde$, respectively) lead to states with well-defined quantum numbers \(J^{PC}\) ~\cite{Maiani:2014aja} (in which total spin $s = J$ is indicated by an outer subscript on the ket):
\begin{alignat}{3}\label{eq:defaultbasis}
	J^{PC}=0^{++}&: & X_{0}&\equiv\ket{0_{\de},0_{\bde}}_{0},\ \ X_{0}'\equiv\ket{1_{\de},1_{\bde}}_{0},\notag\\
	J^{PC}=1^{++}&: & X_{1}&\equiv\frac{1}{\sqrt{2}}\left(\ket{1_{\de},0_{\bde}}_{1}+\ket{0_{\de},1_{\bde}}_{1}\right),\notag\\
	J^{PC}=1^{+-}&: & Z_{\phantom{0}}&\equiv\frac{1}{\sqrt{2}}\left(\ket{1_{\de},0_{\bde}}_{1}-\ket{0_{\de},1_{\bde}}_{1}\right),\notag\\
    & &  \quad Z'&\equiv\ket{1_{\de},1_{\bde}}_{1},\notag\\
	J^{PC}=2^{++}&: & X_{2}&\equiv\ket{1_{\de},1_{\bde}}_{2}.
\end{alignat}

Thus, prior to adding isospin, the \(\Sigma_{g}^{+}\left(1S\right)\) multiplet consists of 6 states.  Including both \(I=0, \, 1\) states increases the total to 12.

For our initial analysis, we ignore possible spin- and isospin-dependent fine structure introduced directly between the fundamental $\de\bde$ pair, and focus upon the possibility that fine structure in the $\Sigma^+_g(1S)$ multiplet arises entirely from distinct couplings to dimeson threshold states ({\it e.g.}, the $D^*$-$D$ mass difference).  That is, in this paper we first introduce the {\it diabatic} dynamical diquark model before incorporating key features of the {\it adiabatic} dynamical diquark model.

\subsection{The Diabatic Model}

The \textit{diabatic} dynamical diquark model~\cite{Lebed:2022vks} extends the Born-Oppenheimer (BO) approach to allow interactions between different configurations of the constituents in a state.  More specifically, this model allows the $\de\bde$ pair \((qQ)(\bar{q}'\bar{Q})\) to mix with dimeson states \(M_1\bar M_2 \equiv (Q\bar{q}')(\bar{Q}q)\), since open heavy-flavor mesons are expected to dominate such couplings~\cite{Lebed:2022vks}.  The formalism employed here was initially developed in the context of molecular spectroscopy~\cite{Baer:2006}, and was first applied to hadronic systems (specifically, heavy quarkonium) in Ref.~\cite{Bruschini:2020voj}.

The tetraquark system of interest contains two heavy color sources that interact with light degrees of freedom (d.o.f.).  Thus, one may appeal to the scale separation $m_Q \gg \Lambda_{\rm QCD}$ in order to develop a nonrelativistic Hamiltonian:
\begin{equation} \label{eq:Ham}
    H=K_{\mathrm{heavy}}+H_{\mathrm{light}}=\frac{\vv{p}^{2}}{2\mu_{\mathrm{heavy}}}+H_{\mathrm{light}},
\end{equation}
where the light field refers to the glue field between the components of a $\de\bde$ state, or glue combined with the light-quark fields for the case of dimeson states.  The adiabatic BO approximation applied to the solution yields
\begin{equation}
    \ket{\psi}=\sum_{i}\int\rmd\vv{r}\ \tilde{\psi}_{i}\left(\vv{r}\right)\ket{\vv{r}}\ket{\xi_{i} \! \left(\vv{r}\right)},
\end{equation}
in which \(\left\{\ket{\xi_{i} \! \left(\vv{r}\right)}\right\}\) forms a complete, orthonormal basis for the light d.o.f.'s\@.  The diabatic formalism introduces a new free parameter \(\vv{r}_{0}\), in terms of which:
\begin{equation}\label{eq:diabaticwavefunctionexpansion}
    \ket{\psi}=\sum_{i}\int \rmd\vv{r}'\tilde{\psi}\left(\vv{r}',\vv{r}_{0}\right)\ket{\vv{r}'}\ket{\xi_{i}\left(\vv{r}_{0}\right)}.
\end{equation}
Note that \(\left\{\ket{\xi_{i}\left(\vv{r}\right)}\right\}\) is an orthonormal basis for any \(\vv{r}\), and in particular for any \(\vv{r}_{0}\).  Consequently, \(\left\{\ket{\xi_{i} \! \left(\vv{r}_{0}\right)}\right\}\) remains an orthonormal basis in the diabatic formalism.  Forming inner products of the Hamiltonian in Eq.~\eqref{eq:Ham} with \(\bra{\xi_{j}\left(\vv{r}_{0}\right)}\) and $\left| \psi \right>$ from Eq.~\eqref{eq:diabaticwavefunctionexpansion} gives
\begin{equation} \label{eq:DiabaticV}
    \sum_{i}\left[-\frac{\hbar^{2}}{2\mu_{i}}\delta_{ji}\grad^{2} \! +V_{ji}\left(\vv{r},\vv{r}_{0}\right)-E\delta_{ji}\right]\tilde{\psi}_{i}\left(\vv{r},\vv{r}_{0}\right) = 0 \, .
\end{equation}

The new coupled-channel potential \(V_{ji}\) in Eq.~(\ref{eq:DiabaticV}) is known as the \textit{diabatic potential matrix}, and is defined as
\begin{equation}
    V_{ji}\left(\vv{r},\vv{r}_{0}\right)\equiv\braket{\xi_{j}\left(\vv{r}_{0}\right)|H_{\mathrm{light}}\left(\vv{r}\right)|\xi_{i}\left(\vv{r}_{0}\right)}.
\end{equation}
We can choose \(\vv{r}_{0}\) to lie far from any potential-energy level crossings, so that \(\ket{\xi_{i}(\vv{r}_{0})}\) may be identified with pure, unmixed configurations that are eigenstates of a single potential-energy component.  The diagonal element $V_{11}$ in this model is the static light-field energy \(V_{\de\bde}\) for a pure \(\de\bde\) state,  determined by lattice QCD~\cite{Berwein:2015vca}:
\begin{equation}\label{eq:diquarkpotential}
    V_{\de\bde}\left(r\right)=-\frac{\alpha}{r}+\sigma r+V_{0}+m_{\de}+m_{\bde},
\end{equation}
with \(\alpha=0.053\)~GeV\(\,\cdot\)\,fm, \(\sigma=1.097\)~GeV\(/\)fm, \(V_{0}=-0.180\)~GeV, and \(m_{\de}=m_{\bde}\,\) still to be determined.  The remaining diagonal elements $V_{ii}$ consist of dimeson threshold potentials \(V^{(i)}_{M_{1}\bar{M}_{2}}\), \(i=1,2,\ldots,N\), set equal to the sum of their masses, thus indicating no interaction between them:
\begin{equation}
    V^{(i)}_{M_{1}\bar{M}_{2}}\left(\vv{r}\right)\to T_{M_{1}\bar{M}_{2}}=M_{1}+M_{2}.
\end{equation}
This assumption is not essential to the model; in fact, nontrivial interaction potentials between $M_1$ and $\bar M_2$ are straightforward to include~\cite{Lebed:2025xbz}.  Off-diagonal elements consist of mixing potentials between the pure diquark-antidiquark (\(\de\bde\)) and pure dimeson (\(M_1 \bar M_2\)) states, written as \(V_{\mathrm{mix}}^{(i)}\left(\vv{r}\right)\).  Direct mixings between the distinct dimeson states are taken to be negligible.  Collecting the terms, the potential is written explicitly as
\begin{equation}\ 
    \vv{V} \! \left(\vv{r}\right) \! = \!\!
    \renewcommand{\arraystretch}{1.5}
    \begin{pmatrix}
        V_{\de\bde}\left(\vv{r}\right) & V_{\mathrm{mix}}^{(1)}\left(\vv{r}\right) & \cdots & V_{\mathrm{mix}}^{(N)}\left(\vv{r}\right)\\
        V_{\mathrm{mix}}^{(1)}\left(\vv{r}\right) & V^{(1)}_{M_{1}\bar{M}_{2}}\left(\vv{r}\right) & & \\
        \vdots & & \ddots  & \\
        V_{\mathrm{mix}}^{(N)}\left(\vv{r}\right) & & & V^{(N)}_{M_{1}\bar{M}_{2}}\left(\vv{r}\right)
    \end{pmatrix}, \label{eq:MixingPotential}
\end{equation}
where blank entries are zero.  The mixing potentials themselves are modeled using a Gaussian form~\cite{Bruschini:2020voj}:
\begin{equation}\label{eq:Gaussianmix}
    |V_{\mathrm{mix}}^{(i)}\left(\vv{r}\right)| \! =\frac{\Delta}{2}\exp\left\{-\frac{1}{2}\frac{\left[V_{\de\bde}\left(r\right)-T^{(i)}_{M_{1}\bar{M_{2}}}\right]^{2}}{\left(\rho\sigma\right)^{2}}\right\},
\end{equation}
where \(\Delta\) indicates the strength of the mixing, \(\sigma\) is the string-tension parameter, and \(\rho\) is the (dimensionless) radial scale associated with the level crossing.  Both \(\Delta\) and \(\rho\) are free parameters to be determined, although \(\sigma\) is the same string tension that appears in Eq.~\eqref{eq:diquarkpotential}.  The precise functional form of Eq.~\eqref{eq:Gaussianmix} is chosen purely for convenience: Its only purpose is to decouple $V_{\de\bde}$ and $V^{(i)}_{M_1 \bar M_2}$ far from threshold.

\subsection{Spin-dependent diabatic potential matrices}\label{sec:spindependentmatrices}

In previous work, the diabatic potential matrix Eq.~\eqref{eq:MixingPotential} was taken to have universal \(\de\bde\)-dimeson mixing potentials.  In general, one expects unique couplings between $\de\bde$ and each dimeson state.  However, heavy-quark spin symmetry constrains the relative mixing amplitudes of different dimeson spin configurations, such as $D\bar D$, $D \bar D^*$, and $D^* \bar D^*$ [generically, $D^{(*)} \bar D^{(*)}$] for the charmed case.  The version of this problem in which the underlying states are $Q\bar Q$ was first addressed in Ref.~\cite{Bruschini:2023zkb}.  Since in the present case the available \(Q\bar{Q}q\bar{q}'\) spins can be coupled to both \(\de\bde\) \([(Qq)(\bar{Q}\bar{q}')]\) and dimeson \([(Q\bar{q}')(\bar{Q}q)]\) configurations, novel applications of Fierz rearrangement between these two coupling schemes are derived in Ref.~\cite{Lebed:2025eqm}, and are then used to determine the group-theoretical factors multiplying a universal mixing function \(h(r)\) for dimeson states related by heavy-quark spin symmetry.

In the \(c\bar{c}q\bar{q}'\) sector, where \(D^{(*)}\) mesons provide the thresholds for \(\Sigma_{g}^{+}\), energies are expressed with reference to the $D^{(*)}$ spin average:
\begin{equation} \label{eq:DDspinavemass}
    m_{0}\equiv\frac{1}{4}\left(m_{D}+3m_{D^{\ast}}\right) .
\end{equation}
The (fine structure) difference in mass between the ground and excited $D^{(*)}$ mesons is
\begin{equation} \label{eq:Dfine}
    m_{\Delta}\equiv m_{D^{\ast}}-m_{D},
\end{equation}
and so the various dimeson thresholds can be written as \(-\frac{3}{2}m_{\Delta}\), \(-\frac{1}{2}m_{\Delta}\), and \(+\frac{1}{2}m_{\Delta}\) with respect to $m_0$, respectively, for $D \bar D$, $D \bar D^*$, $D^* \bar D^*$.  We take the universal mixing function $h(r)$ to be the Gaussian defined in Eq.~\eqref{eq:Gaussianmix}.  The rigorous derivation of the spin-dependent mixing potentials for each $J^{PC}$, including all allowed partial waves, is given in Ref.~\cite{Lebed:2025eqm}.  The explicit spin-dependent diabatic matrices are listed in Section III of this reference.

The $D^{(*)}\bar D^{(*)}$ combinations ${\cal D}_J$ that couple to each channel are also defined in Ref.~\cite{Lebed:2025eqm} [Eqs.~(28)-(29)]; for our present purposes, their only relevant properties are that ${\cal D}_J$ carries total angular momentum $J$, and that the number of asterisks on ${\cal D}_J$ indicates how many of the $D^{(*)}\bar D^{(*)}$ pair are actually $D^*$'s.  The superscripts $g$ and $u$ on ${\cal D}_J$ indicate specific symmetry properties.

Restricting to the \(S\)-wave channels of the multiplet, the potentials of interest are
\begin{equation}\label{eq:3872diabaticmatrix}
    \vv{V}^{1^{++}} \! \left(r\right)=
    \renewcommand{\arraystretch}{1.5}   
    \begin{pmatrix}
        V_{\de\bde}\left(r\right) & +\frac{1}{3}h\!\left(r\right)\\
        +\frac{1}{3}h\!\left(r\right) & -\frac{1}{2}m_{\Delta}
    \end{pmatrix},
\end{equation}
coupling the \(\de\bde\) state to the \(D\bar{D}^{\ast}\) combination called \(\mcD_{1}^{\ast g}\), and
\begin{equation}\label{eq:39004020diabaticmatrix}
    \vv{V}^{1^{+-}} \! \left(r\right)=
    \renewcommand{\arraystretch}{1.5}   
    \begin{pmatrix}
        V_{\de\bde}\left(r\right) & -\frac{i}{3}h\!\left(r\right) & 0 & 0\\
        +\frac{i}{3}h\!\left(r\right) & -\frac{1}{2}m_{\Delta} & 0 & 0\\
        0 & 0 & V_{\de\bde}\left(r\right) & +\frac{i}{3}h\!\left(r\right)\\
        0 & 0 & -\frac{i}{3}h\!\left(r\right) & +\frac{1}{2}m_{\Delta}
    \end{pmatrix},
\end{equation}
corresponding, in order, to coupling between the states \(\left\{Z',\mcD_{1}^{\ast u},Z,\mcD_{1}^{\ast\ast}\right\}\).  The matrix \(\vv{V}^{1^{++}}\) in this work only couples to the state \(\chi_{c1}(3872)\), while \(\vv{V}^{1^{+-}}\) couples to both \(T_{c\bar{c}1}(3900)\) and \(T_{c\bar{c}1}(4020)\).  For numerical calculations, it is convenient to treat each member of the multiplet as its own separate coupled-channel problem (indeed, the observed mass eigenvalues are well separated).  To distinguish \(T_{c\bar{c}1}(3900)\) and \(T_{c\bar{c}1}(4020)\), we employ a mixing angle \(\theta\in [0,\pi/2]\) to be determined later, thus defining a new basis:
\begin{equation}\label{eq:zetarotatedbasis}
    \begin{pmatrix}
		\zeta \\ \zeta'
	\end{pmatrix}\equiv\vv{R}\left(\theta\right)\begin{pmatrix}
		Z\\Z'
	\end{pmatrix},
\end{equation}
where \(\vv{R}\left(\theta\right)\) is the usual rotation matrix,
\begin{equation}\label{eq:rotationmatrix}
    \vv{R}\left(\theta\right)\equiv\begin{pmatrix}
		\cos\theta & -\sin\theta\\
		\sin\theta & \cos\theta
	\end{pmatrix}.
\end{equation}
This separation effectively splits the \(4\times 4\) matrix  $\vv{V}^{1^{+-}}$ into a pair of \(3\times 3\) matrices,
%
%\begin{widetext}
\begin{align}\label{eq:split39004020diabaticmatrices}
    \vv{V}_{\zeta}^{1^{+-}}\!\!\left(r\right)&\equiv\!
    \renewcommand{\arraystretch}{1.5}
    \begin{pmatrix}
        V_{\de\bde}\left(r\right) & +\frac{i}{3}\sin \theta \, h\!\left(r\right) & +\frac{i}{3}\cos \theta \, h\!\left(r\right)\\[4pt]
        -\frac{i}{3}\sin \theta \, h\!\left(r\right) & -\frac{1}{2}m_{\Delta} & 0\\[4pt]
        -\frac{i}{3}\cos \theta \, h\!\left(r\right) & 0 & +\frac{1}{2}m_{\Delta}
    \end{pmatrix} \! , \notag\\
    \vv{V}_{\zeta'}^{1^{+-}}\!\!\left(r\right)&\equiv\!
    \renewcommand{\arraystretch}{1.5}
    \begin{pmatrix}
        V_{\de\bde}\left(r\right) & -\frac{i}{3}\cos \theta \, h\!\left(r\right) & +\frac{i}{3}\sin \theta \, h\!\left(r\right)\\[4pt]
        +\frac{i}{3}\cos \theta \, h\!\left(r\right) & -\frac{1}{2}m_{\Delta} & 0\\[4pt]
        -\frac{i}{3}\sin \theta \, h\!\left(r\right) & 0 & +\frac{1}{2}m_{\Delta}
    \end{pmatrix} \! ,
\end{align}
%\end{widetext}
%
with basis orderings \(\left\{\zeta,\mcD_{1}^{\ast u}, \mcD_{1}^{\ast\ast}\right\}\) and \(\left\{\zeta',\mcD_{1}^{\ast u}, \mcD_{1}^{\ast\ast}\right\}\), respectively.

\subsection{Numerical Results}

The diabatic potential matrices depend upon four parameters that remain numerically undetermined at this stage: \(\{\theta,\Delta,\rho,m_{\de}\}\).  For any choice of these parameters, Eqs.~\eqref{eq:3872diabaticmatrix} and \eqref{eq:split39004020diabaticmatrices} define the potentials, which are inserted into the coupled-channel Schr\"{o}dinger equation and are solved numerically using the renormalized Numerov method~\cite{Johnson:1978}.  The scattering matrix \(S\) thus obtained is used to extract the partial-wave scattering amplitudes~\cite{Bruschini:2021ckr,Lebed:2023kbm},
\begin{equation}
    f^{i\gets i'}_{J^{PC}}=\frac{(S-\Id)_{ii'}}{2ip^{(i)}}
\end{equation}
which allows one to calculate a normalized, dimensionless cross section \(\bar{\sigma}\)~\cite{Bruschini:2021ckr,Lebed:2023kbm},
\begin{equation}
\label{eq:normcrosssection}
    \bar{\sigma}^{i\gets i'}_{J^{PC}}\equiv\sum_{k,k'}\left|p^{(i)}f^{i\gets i'}_{J^{PC};k,k'}\right|^{2} ,
\end{equation}
where $p^{(i)} \equiv \sqrt{2\mu^{(i)} (E - T^{(i)}})$ is the relative momentum of the dimeson pair, and $k,k^\prime$ indicate quantum numbers used to denote specific partial waves.  Resonances may be identified from Breit-Wigner-like structures in \(\bar{\sigma}\).  The objective is therefore to determine whether a common set \(\{\theta,\Delta,\rho,m_{\de}\}\) exists for which structures associated with the \(\chi_{c1}(3872)\), \(T_{c\bar{c}1}(3900)\), and \(T_{c\bar{c}1}(4020)\) simultaneously occur at their experimental masses reported by the PDG in Eq.~\eqref{eq:PDGmasses}.  Decay widths were not used as constraints in this analysis.

Due to the four-dimensional parameter space and computational cost of each numerical solution, the search was performed in stages.  An initial broad scan identified parameter sets for which resonances appeared in all three systems within a 200~MeV energy window centered on their experimental masses.  These points were retained as candidate solutions, and were subsequently inspected to determine whether they produced agreement with the three observed masses.  Any promising candidates would then be used as the center of a localized search in order to refine the numerical values.  Latin hypercube sampling was used to generate the initial broad samples over the ranges:
\begin{align}\label{eq:parameterbounds}
    \theta&\in[\,0,\pi/2\,]\ \mathrm{rad},\notag\\
    \Delta&\in[\,0.01,1.5\,]\ \mathrm{GeV},\notag\\
    \rho&\in[\,0.01,1.5\,]\ \mathrm{fm},\notag\\
    m_{\de}&\in[\,1.8,2.1\,]\ \mathrm{GeV}.
\end{align}
The adopted ranges for $\Delta$ and $\rho$ were simply chosen to reflect typical hadronic scales.

To estimate the sampling density required for the parameter scan, we performed preliminary trial calculations by varying each parameter independently.  Changes on the order of 0.05 in the respective units of each parameter were found to produce appreciable changes in the resulting cross sections.  Using this criterion to obtain an approximate characteristic spacing over the parameter ranges produces roughly 167,000 samples required to explore the phenomenologically relevant region.  The actual scan was generously expanded to 250,000 points in order to provide denser coverage.  Of this sample, only 358 simulations returned resonances in all three systems within the energy window.  Manual inspection of the remaining candidates revealed that none of them were able to simultaneously reproduce the observed spectrum particularly well; the three states of interest generally clustered much closer to 4~GeV than to the $\simeq 3.9$~GeV masses of the experimentally observed states.

Nevertheless, a representative best (more accurately, ``least bad'') candidate was selected by minimizing the total absolute error of the three calculated resonance positions with respect to their corresponding experimental masses.  Although this parameter set provides the closest simultaneous agreement among the sampled points, substantial discrepancies remain in the mass splitting.  Its parameters were found to be:
\begin{align}\label{eq:bestdiabaticparameters}
    \theta&=0.624\ \mathrm{rad},\notag\\
    \Delta&=0.132\ \mathrm{GeV},\notag\\
    \rho&=0.659\ \mathrm{fm},\notag\\
    m_{\de}&=1.822\ \mathrm{GeV}.
\end{align}
% Was: 0.6237, 0.1317, 0.6594, 1.822
%
These values result in predicted masses of 3.9815~GeV, 3.9805~GeV, and 4.0250~GeV for \(\chi_{c1}(3872)\), \(T_{c\bar{c}1}(3900)\), and \(T_{c\bar{c}1}(4020)\), respectively.  No small perturbations in \(\{\theta,\Delta,\rho,m_{\de}\}\) could bring the resonances, displaced by several hundred MeV, closer to their experimental masses [apart from \(T_{c\bar{c}1}(4020)\), which was already nearby].  Perhaps even worse, many of these candidates (including this one) do not preserve the observed mass ordering, often predicting that the mass of \(T_{c\bar{c}1}(3900)\) becomes slightly smaller than that of \(\chi_{c1}(3872)\).  Such results are thus considered as strong evidence that the diabatic dynamical diquark model is incapable of reproducing observed experimental values on its own, using only couplings to $D^{(*)}\bar D^{(*)}$ dimeson threshold states whose relative couplings are determined using heavy-quark spin symmetry.  The intrinsic spin (and isospin) couplings of the fundamental $\de\bde$ pair thus appear to be absolutely essential for reproducing the correct experimental mass eigenvalues for the states.

\section{The Adiabatic Correction}\label{sec:AdiabaticCorrection}

\subsection{Formalism}

The diabatic dynamical diquark model historically was preceded by its original {\it adiabatic} form~\cite{Lebed:2017min}, in which the multiplets of exotic hadrons are still determined by BO potentials, but couplings to nearby di-hadron thresholds are ignored.  In the model's adiabatic form, the fine structure of each multiplet is determined by interactions between the $\de$ and $\bde$ quasiparticles and how they are assembled.  In order to calculate these effects efficiently, one may rewrite the basis for the $\Sigma^+_g$ multiplet Eq.~\eqref{eq:defaultbasis}, expressed using diquark spins $s_\de$, $s_\bde$, in a basis of good spin for heavy-quark ($s_{Q\bar Q}$) and light-quark ($s_{q\bar q}$) pairs:
\begin{alignat}{3}\label{eq:tildebasis}
	\tilde{X}_{0}&\equiv&\Ket{0_{q\bar{q}'},0_{Q\bar{Q}}}_{0}&=+\frac{1}{2}X_{0}+\frac{\sqrt{3}}{2}X_{0}',\notag\\
    \tilde{X}_{0}'&\equiv&\Ket{1_{q\bar{q}'},1_{Q\bar{Q}}}_{0}&=+\frac{\sqrt{3}}{2}X_{0}-\frac{1}{2}X_{0}',\notag\\
    \tilde{Z}&\equiv&\Ket{1_{q\bar{q}'},0_{Q\bar{Q}}}_{1}&=\frac{1}{\sqrt{2}}\left(Z'+Z\right),\notag\\
    \tilde{Z}'&\equiv&\Ket{0_{q\bar{q}'},1_{Q\bar{Q}}}_{1}&=\frac{1}{\sqrt{2}}\left(Z'-Z\right),\notag\\
    X_{1}&=&\Ket{1_{q\bar{q}'},1_{Q\bar{Q}}}_{1} &,\notag\\
    X_{2}&=&\Ket{1_{q\bar{q}'},1_{Q\bar{Q}}}_{2}&.
\end{alignat}
For \(Q\bar{Q}q\bar{q}'\) states, the Hamiltonian is taken to be~\cite{Giron:2019cfc,Giron:2021sla}
\begin{align}\label{eq:adiabaticspinhamiltonian}
    H&=M_{0}+\Delta H_{\kappa_{qQ}}+\Delta H_{V_{0}}+\Delta H_{V_{8}}\notag\\
    &=M_{0}+2\kappa_{qQ}\left(\vv{s}_{q}\cdot\vv{s}_{Q}+\vv{s}_{\bar{q}}\cdot\vv{s}_{\bar{Q}}\right)\notag\\
    &\phantom{=}+V_{0}\left(\vv{\tau}_{q}\cdot\vv{\tau}_{\bar{q}}\right)\left(\vv{\sigma}_{q}\cdot\vv{\sigma}_{\bar{q}}\right)%\notag\\
   %&\phantom{=}
    +V_{8}\left(\lambda_{a}^{8}\vv{\sigma}_{q}\right)\cdot\left(\lambda_{a}^{8}\vv{\sigma}_{\bar{q}}\right),
\end{align}
where \(M_{0}\) is the common multiplet mass, \(\Delta H_{\kappa_{qQ}}\) is the spin-spin interaction {\it within each} diquark, \(\Delta H_{V_{0}}\) is an isospin-spin interaction of strength \(V_{0}\) {\it between} the light quarks $q$ in $\de$ and $\bar q^\prime$ in $\bde$, analogous to pion exchange, and \(\Delta H_{V_{8}}\) is an interaction modeled on \(\eta\)-nucleon coupling.  We assume that both \(q,q'\) give the same coupling \(\kappa_{qQ}\).  In this work, we neglect the \(\eta\)-like coupling, and hence take \(V_{8}=0\).  In the basis of Eq.~\eqref{eq:tildebasis}, using boldface for matrices (with elements arranged in the order \(s_{Q\bar{Q}}=0,1\)), the matrix elements for all 12 states of $\Sigma^+_g(1S)$ read:
\begin{align}\label{eq:adiabaticmasssplittingcorr}
    \tilde{\vv{M}}_{0^{++}}^{I=0}&=M_{0}\vv{I}-\kappa_{qQ}\begin{pmatrix}
		0 & \sqrt{3}\\
		\sqrt{3} & 2
	\end{pmatrix}-3V_{0}\begin{pmatrix}
		-3 & 0\\
		0 & 1
	\end{pmatrix},\notag\\
	\tilde{\vv{M}}_{0^{++}}^{I=1}&=M_{0}\vv{I}-\kappa_{qQ}\begin{pmatrix}
		0 & \sqrt{3}\\
		\sqrt{3} & 2
	\end{pmatrix}+V_{0}\begin{pmatrix}
		-3 & 0\\
		0 & 1
	\end{pmatrix},\notag\\
	\tilde{\vv{M}}_{1^{+-}}^{I=0}&=M_{0}\vv{I}+\kappa_{qQ}\begin{pmatrix}
		0 & 1\\
		1 & 0
	\end{pmatrix}-3V_{0}\begin{pmatrix}
		1 & 0\\
		0 & -3
	\end{pmatrix},\notag\\
	\tilde{\vv{M}}_{1^{+-}}^{I=1}&=M_{0}\vv{I}+\kappa_{qQ}\begin{pmatrix}
		0 & 1\\
		1 & 0
	\end{pmatrix}+V_{0}\begin{pmatrix}
		1 & 0\\
		0 & -3
	\end{pmatrix},\notag\\
	M_{1^{++}}^{I=0}&=M_{0}-\kappa_{qQ}-3V_{0},\notag\\
	M_{1^{++}}^{I=1}&=M_{0}-\kappa_{qQ}+V_{0},\notag\\
	M_{2^{++}}^{I=0}&=M_{0}+\kappa_{qQ}-3V_{0},\notag\\
	M_{2^{++}}^{I=1}&=M_{0}+\kappa_{qQ}+V_{0}.
\end{align}

From this point, the parameters \(M_{0}\), \(\kappa_{qQ}\), and \(V_{0}\) may be extracted uniquely from the PDG masses in Eq.~\eqref{eq:PDGmasses} for $\chi_{c1}(3872)$, $T_{c{\bar c}1}(3900)$, and $T_{c{\bar c}1}(4020)$ and used to calculate masses for all other states in the multiplet, as originally done in Ref.~\cite{Giron:2021sla}.  The resulting spectrum describes the fine structure of the $c\bar c q\bar q^\prime$ $\Sigma^+_g(1S)$ multiplet arising from internal (iso)spin interactions.  Within the adiabatic treatment, however, these levels remain discrete bound eigenstates due to the confining \(\de\bde\) potential Eq.~\eqref{eq:diquarkpotential}.  Even when a predicted mass lies above an open dimeson threshold, no decay width is generated since the corresponding continuum channel is not included in the bound-state problem.

The diabatic dynamical diquark model~\cite{Lebed:2022vks} was developed precisely to address this limitation, by coupling \(\de\bde\) to the appropriate dimeson channel(s).  An energy eigenstate with mass lying above an open threshold may consequently appear as a resonance rather than as a discrete bound state, but in both cases the mass and width are modified by the threshold interaction.  In addition, the spin-dependent diabatic potential introduced in Eq.~\eqref{eq:MixingPotential} further distinguishes the allowed dimeson spin configurations and their relative couplings, as seen in Eqs.~\eqref{eq:3872diabaticmatrix} and \eqref{eq:split39004020diabaticmatrices}. However, \(V_{\de\bde}\left(r\right)\) as used in Sec.~\ref{sec:spindependentmatrices} remains common to all members of the multiplet, and hence does not contain the fine-structure splittings predicted by the full spin-dependent adiabatic Hamiltonian, Eq.~\eqref{eq:adiabaticspinhamiltonian}.

These two descriptions, as thus far described, independently attempt to explain the fine structure of the $c\bar c q\bar q^\prime$ multiplet $\Sigma^+_g(1S)$.  The original adiabatic model determines the spin-dependent fine structure of the multiplet, and matches the mass eigenvalues of the known states perfectly, but does not allow for couplings to the dimeson states that clearly influence their structure.  On the other hand, the diabatic model determines how the $\de\bde$ configurations are modified by, and couple to, the open dimeson continuum, but we have seen that their distinct couplings alone are insufficient for explaining the spectrum of the known states $\chi_{c1}(3872)$, $T_{c{\bar c}1}(3900)$, and $T_{c{\bar c}1}(4020)$.  Combining the two descriptions allows the spin-dependent structure of the multiplet to be retained, while treating the corresponding states within a coupled-channel scattering framework.

Since the adiabatic description alone is already sufficient to reproduce the known mass eigenvalues, our approach here is to start with the diabatic calculations described in Sec.~\ref{sec:spindependentmatrices}, for which no reasonable choice of parameters \(\{\theta,\Delta,\rho,m_{\de}\}\) is sufficient to describe the known spectrum, and find the {\em minimal values\/} of the $\de\bde$ spin-dependent Hamiltonian parameters \(\{\kappa_{qQ}, V_0 \}\) required to reproduce the spectrum.

In order to incorporate the spin- and isospin-dependent effects of the adiabatic model into the diabatic model, the multiplet average mass \(M_{0}\) in Eq.~\eqref{eq:adiabaticmasssplittingcorr} must be re-interpreted relative to the value used in the pure diabatic calculation.  First, the zero of the mass scale in the pure diabatic calculation is set by the $D$-$D^*$ spin-averaged mass $m_0$~[Eq.~\eqref{eq:DDspinavemass}].  Second, the diabatic calculation of Sec.~\ref{sec:spindependentmatrices} introduces its own spin-dependent corrections [Eq.~\eqref{eq:Dfine}] to the mass spectrum.  We therefore introduce a parameter $\Delta M_0$ into the $\de\bde$ potential that serves purely as a mass shift of an amount common to all $\de\bde$ states relative to the dimeson levels.  The mass relations of Eq.~\eqref{eq:adiabaticmasssplittingcorr} are then re-written as:
\begin{align}\label{eq:adiabaticmasssplittingtilde}
    \Delta\tilde{\vv{M}}_{0^{++}}^{I=0}&=\Delta M_{0}\vv{I}-\kappa_{qQ}\begin{pmatrix}
		0 & \sqrt{3}\\
		\sqrt{3} & 2
	\end{pmatrix}-3V_{0}\begin{pmatrix}
		-3 & 0\\
		0 & 1
	\end{pmatrix},\notag\\
	\Delta\tilde{\vv{M}}_{0^{++}}^{I=1}&=\Delta M_{0}\vv{I}-\kappa_{qQ}\begin{pmatrix}
		0 & \sqrt{3}\\
		\sqrt{3} & 2
	\end{pmatrix}+V_{0}\begin{pmatrix}
		-3 & 0\\
		0 & 1
	\end{pmatrix},\notag\\
	\Delta\tilde{\vv{M}}_{1^{+-}}^{I=0}&=\Delta M_{0}\vv{I}+\kappa_{qQ}\begin{pmatrix}
		0 & 1\\
		1 & 0
	\end{pmatrix}-3V_{0}\begin{pmatrix}
		1 & 0\\
		0 & -3
	\end{pmatrix},\notag\\
	\Delta\tilde{\vv{M}}_{1^{+-}}^{I=1}&=\Delta M_{0}\vv{I}+\kappa_{qQ}\begin{pmatrix}
		0 & 1\\
		1 & 0
	\end{pmatrix}+V_{0}\begin{pmatrix}
		1 & 0\\
		0 & -3
	\end{pmatrix},\notag\\
	\Delta M_{1^{++}}^{I=0}&=\Delta M_{0}-\kappa_{qQ}-3V_{0},\notag\\
	\Delta M_{1^{++}}^{I=1}&=\Delta M_{0}-\kappa_{qQ}+V_{0},\notag\\
	\Delta M_{2^{++}}^{I=0}&=\Delta M_{0}+\kappa_{qQ}-3V_{0},\notag\\
	\Delta M_{2^{++}}^{I=1}&=\Delta M_{0}+\kappa_{qQ}+V_{0}.
\end{align}

Additionally, we introduce an independent additive shift \(C\) to $V_{\de\bde}(r)$ in the 11 element of each diabatic potential matrix in Eqs.~\eqref{eq:3872diabaticmatrix}-\eqref{eq:split39004020diabaticmatrices}.  These three \(C\) values are then unpacked to produce values for \(\{\Delta M_0, \kappa_{qQ}, V_0 \}\).

\subsection{Incorporation into Diabatic Potentials}

As done for the \(J^{PC}=1^{+-}\) system in Eq.~\eqref{eq:39004020diabaticmatrix}, the \(J^{PC}=0^{++}\) matrix must also be expressed using a mixing angle, in order to make subsequent numerical calculations more convenient.  Furthermore, the corrections from Eq.~\eqref{eq:adiabaticmasssplittingtilde} must be rotated into the new basis [Eq.~\eqref{eq:defaultbasis} to Eq.~\eqref{eq:tildebasis}].  We define these basis transformations via:
\begin{align}
    \begin{pmatrix}
		\tilde{X}_{0}\\ \tilde{X}_{0}'
	\end{pmatrix}&=\vv{U}_{0^{++}}\begin{pmatrix}
		X_{0}\\ X_{0}'
	\end{pmatrix},
	& \begin{pmatrix}
		\tilde{Z}\\ \tilde{Z}'
	\end{pmatrix}&=\vv{U}_{1^{+-}}\begin{pmatrix}
		Z\\ Z'
	\end{pmatrix}.
\end{align}
where
\begin{align}
\label{eq:Umatrices}
 \vv{U}_{0^{++}}&\equiv\begin{pmatrix}
		+\frac{1}{2} & \frac{\sqrt{3}}{2}\\[4pt]
		\frac{\sqrt{3}}{2} & -\frac{1}{2}
	\end{pmatrix}, & \vv{U}_{1^{+-}}&\equiv\frac{1}{\sqrt{2}}\begin{pmatrix}
		\phantom{-}1 & 1\\
		-1 & 1
	\end{pmatrix}.
\end{align}
In parallel to Eq.~\eqref{eq:zetarotatedbasis}, we now define
\begin{equation}\label{eq:xi0rotatedbasis}
    \begin{pmatrix}
		\xi_{0}\\ \xi_{0}'
	\end{pmatrix}\equiv\vv{R}\left(\theta\right)\begin{pmatrix}
		X_{0}\\ X_{0}'
	\end{pmatrix}.
\end{equation}
Note that \(\xi_{0}\) here is unrelated to the complete set of eigenstates \(\left\{\ket{\xi_{i}\left(\vv{r}\right)}\right\}\) in Eq.~\eqref{eq:diabaticwavefunctionexpansion}.  The sectors \(0^{++}\) and \(1^{+-}\) both contain two \(\de\bde\) configurations, which we separate using their respective mixing angles (both labeled as $\theta$ for convenience).  This expansion results in four general diabatic potentials parametrized by their respective mixing angle \(\theta\) and $V_{11}$ correction \(C\), which then holds parametrically for either isospin sector (but assumes differing numerical values for $I=0$ and $I=1$):
\begin{widetext}
    \begin{align}
        \vv{V}_{\xi_{0}}^{0^{++}}\left(r,\theta,C\right)&\equiv
        \renewcommand{\arraystretch}{1.5}       
        \begin{pmatrix}
    		V_{\de\bde}\left(r\right)+C & -\left(\frac{1}{6}\cos\theta+\frac{1}{2\sqrt{3}}\sin \theta\right)h\!\left(r\right) & \left(\frac{1}{2\sqrt{3}}\cos\theta-\frac{1}{6}\sin \theta\right)h\!\left(r\right)\\[5pt]
    		-\left(\frac{1}{6}\cos\theta+\frac{1}{2\sqrt{3}}\sin \theta\right)h\!\left(r\right) & -\frac{3}{2}m_{\Delta} & 0\\[5pt]
    		\left(\frac{1}{2\sqrt{3}}\cos\theta-\frac{1}{6}\sin \theta\right)h\!\left(r\right) & 0 & +\frac{1}{2}m_{\Delta}
    	\end{pmatrix} , \notag\\
    	\vv{V}_{\xi_{0}'}^{0^{++}}\left(r,\theta,C\right)&\equiv
        \renewcommand{\arraystretch}{1.5}       
        \begin{pmatrix}
    		V_{\de\bde}\left(r\right)+C & \left(-\frac{1}{6}\sin\theta+\frac{1}{2\sqrt{3}}\cos \theta\right)h\!\left(r\right) & \left(\frac{1}{2\sqrt{3}}\sin\theta+\frac{1}{6}\cos \theta\right)h\!\left(r\right)\\[5pt]
    		\left(-\frac{1}{6}\sin\theta+\frac{1}{2\sqrt{3}}\cos \theta\right)h\!\left(r\right) & -\frac{3}{2}m_{\Delta} & 0\\[5pt]
    		\left(\frac{1}{2\sqrt{3}}\sin\theta+\frac{1}{6}\cos \theta\right)h\!\left(r\right) & 0 & +\frac{1}{2}m_{\Delta}
    	\end{pmatrix} , \notag\\
    	\vv{V}_{\zeta}^{1^{+-}}\left(r,\theta,C\right)&\equiv
        \renewcommand{\arraystretch}{1.5}
        \begin{pmatrix}
    		V_{\de\bde}\left(r\right)+C & +\frac{i}{3}\sin \theta \, h\!\left(r\right) & +\frac{i}{3}\cos \theta \, h\!\left(r\right)\\[4pt]
    		-\frac{i}{3}\sin \theta \, h\!\left(r\right) & -\frac{1}{2}m_{\Delta} & 0\\[4pt]
    		-\frac{i}{3}\cos \theta \, h\!\left(r\right) & 0 & +\frac{1}{2}m_{\Delta}
    	\end{pmatrix} , \notag\\
    	\vv{V}_{\zeta'}^{1^{+-}}\left(r,\theta,C\right)&\equiv
        \renewcommand{\arraystretch}{1.5}       
        \begin{pmatrix}
    		V_{\de\bde}\left(r\right)+C & -\frac{i}{3}\cos \theta \, h\!\left(r\right) & +\frac{i}{3}\sin \theta \, h\!\left(r\right)\\[4pt]
    		+\frac{i}{3}\cos \theta \, h\!\left(r\right) & -\frac{1}{2}m_{\Delta} & 0\\[4pt]
    		-\frac{i}{3}\sin \theta \, h\!\left(r\right) & 0 & +\frac{1}{2}m_{\Delta}
    	\end{pmatrix} .
    \end{align}
\end{widetext}
The masses of all 12 states in the multiplet can then be identified using resonance structures in
\begin{widetext}
    \begin{align}\label{eq:resonancemultiplet}
    	\vv{M}_{0^{++}}^{I=0}&=\begin{cases}
    		\vv{V}_{\xi_{0}}^{0^{++}}\!\left(r,\,\theta_{0^{++}}^{I=0},\,[\vv{H}_{0^{++}}^{I=0}]_{11}\right)\\[8pt]
    		\vv{V}_{\xi_{0}'}^{0^{++}}\!\left(r,\,\theta_{0^{++}}^{I=0},\,[\vv{H}_{0^{++}}^{I=0}]_{22}\right)
    	\end{cases},\qquad\vv{H}_{0^{++}}^{I=0}=\vv{R}\!\left(\theta_{0^{++}}^{I=0}\right)\vv{U}_{0^{++}}^{\dagger} \, \Delta\tilde{\vv{M}}_{0^{++}}^{I=0} \,\, \vv{U}^{\vphantom{\dagger}}_{0^{++}}\,\vv{R}^{\dagger}\!\left(\theta_{0^{++}}^{I=0}\right),\notag\\
    	\vv{M}_{0^{++}}^{I=1}&=\begin{cases}
    		\vv{V}_{\xi_{0}}^{0^{++}}\!\left(r,\,\theta_{0^{++}}^{I=1},\,[\vv{H}_{0^{++}}^{I=1}]_{11}\right)\\[8pt]
    		\vv{V}_{\xi_{0}'}^{0^{++}}\!\left(r,\,\theta_{0^{++}}^{I=1},\,[\vv{H}_{0^{++}}^{I=1}]_{22}\right)
    	\end{cases},\qquad\vv{H}_{0^{++}}^{I=1}=\vv{R}\!\left(\theta_{0^{++}}^{I=1}\right)\vv{U}_{0^{++}}^{\dagger} \,\Delta\tilde{\vv{M}}_{0^{++}}^{I=1} \,\, \vv{U}_{0^{++}}^{\vphantom{\dagger}}\,\vv{R}^{\dagger}\!\left(\theta_{0^{++}}^{I=1}\right),\notag\\
    	\vv{M}_{1^{+-}}^{I=0}&=\begin{cases}
    		\vv{V}_{\zeta}^{1^{+-}}\!\left(r,\,\theta_{1^{+-}}^{I=0},\,[\vv{H}_{1^{+-}}^{I=0}]_{11}\right)\\[8pt]
    		\vv{V}_{\zeta'}^{1^{+-}}\!\left(r,\,\theta_{1^{+-}}^{I=0},\,[\vv{H}_{1^{+-}}^{I=0}]_{22}\right)
    	\end{cases},\qquad\vv{H}_{1^{+-}}^{I=0}=\vv{R}\!\left(\theta_{1^{+-}}^{I=0}\right)\vv{U}_{1^{+-}}^{\dagger}\,\Delta\tilde{\vv{M}}_{1^{+-}}^{I=0}\,\,\vv{U}_{1^{+-}}^{\vphantom{\dagger}}\,\vv{R}^{\dagger}\!\left(\theta_{1^{+-}}^{I=0}\right),\notag\\
    	\vv{M}_{1^{+-}}^{I=1}&=\begin{cases}
    		\vv{V}_{\zeta}^{1^{+-}}\!\left(r,\,\theta_{1^{+-}}^{I=1},\,[\vv{H}_{1^{+-}}^{I=1}]_{11}\right)\\[8pt]
    		\vv{V}_{\zeta'}^{1^{+-}}\!\left(r,\,\theta_{1^{+-}}^{I=1},\,[\vv{H}_{1^{+-}}^{I=1}]_{22}\right)
    	\end{cases},\qquad\vv{H}_{1^{+-}}^{I=1}=\vv{R}\!\left(\theta_{1^{+-}}^{I=1}\right)\vv{U}_{1^{+-}}^{\dagger}\,\Delta\tilde{\vv{M}}_{1^{+-}}^{I=1}\,\,\vv{U}_{1^{+-}}^{\vphantom{\dagger}}\,\vv{R}^{\dagger}\!\left(\theta_{1^{+-}}^{I=1}\right),\notag\\
    	\vv{M}_{1^{++}}^{I=0}&=
        \renewcommand{\arraystretch}{1.5}       
        \begin{pmatrix}
    		V_{\de\bde}\left(r\right)+H_{1^{++}}^{I=0} & +\frac{1}{3}h\!\left(r\right)\\
    		+\frac{1}{3}h\!\left(r\right) & -\frac{1}{2}m_{\Delta}
    	\end{pmatrix} \ \ \ ,\qquad H_{1^{++}}^{I=0}=m\Delta M_{0}-\kappa_{qQ}-3V_{0},\notag\\
    	\vv{M}_{1^{++}}^{I=1}&=
        \renewcommand{\arraystretch}{1.5}       
        \begin{pmatrix}
    		V_{\de\bde}\left(r\right)+H_{1^{++}}^{I=1} & +\frac{1}{3}h\!\left(r\right)\\
    		+\frac{1}{3}h\!\left(r\right) & -\frac{1}{2}m_{\Delta}
    	\end{pmatrix} \ \ \ ,\qquad H_{1^{++}}^{I=1}=\Delta M_{0}-\kappa_{qQ}+V_{0},\notag\\
    	\vv{M}_{2^{++}}^{I=0}&=
        \renewcommand{\arraystretch}{1.5}       
        \begin{pmatrix}
    		V_{\de\bde}\left(r\right)+H_{2^{++}}^{I=0} & +\frac{1}{3}h\!\left(r\right)\\
    		+\frac{1}{3}h\!\left(r\right) & +\frac{1}{2}m_{\Delta}
    	\end{pmatrix} \ \ \ ,\qquad H_{2^{++}}^{I=0}=\Delta M_{0}+\kappa_{qQ}-3V_{0},\notag\\
    	\vv{M}_{2^{++}}^{I=1}&=
        \renewcommand{\arraystretch}{1.5}
        \begin{pmatrix}
    		V_{\de\bde}\left(r\right)+H_{2^{++}}^{I=1} & +\frac{1}{3}h\!\left(r\right)\\
    		+\frac{1}{3}h\!\left(r\right) & +\frac{1}{2}m_{\Delta}
    	\end{pmatrix} \ \ \ ,\qquad H_{2^{++}}^{I=1}=\Delta M_{0}+\kappa_{qQ}+V_{0}.
    \end{align}
\end{widetext}
where \(\left[\vv{H}\right]_{ij}\) refers to the \(i\)th row and \(j\)th column of the respective Hamiltonian.\footnote{We note a typo in the $2^{++}$ case of Ref.~\cite{Lebed:2025eqm} [Eqs.~(73)--(74)]: The upper $2\times 2$ blocks, $\bm{V}^{(g,1)}(r)$ and $\{X_1, {\cal D}^{*g}_1 \}$, respectively, do not satisfy angular momentum constraints and thus should be deleted.}
\vspace{3.25ex}

\subsection{Numerical Results}

In principle, were multiple clear spectroscopic measurements in the $\Sigma^+_g(1S)$ multiplet available, it would be possible to determine simultaneous values for the diabatic parameters \(\{\theta, \, \Delta, \, \rho, \, m_{\de}\}\) and the adiabatic parameters \( \{\Delta M_0, \kappa_{qQ}, V_0\} \).  However, with only \(\chi_{c1}(3872)\), \(T_{c\bar{c}1}(3900)\), and \(T_{c\bar{c}1}(4020)\) masses known, we follow a two-step process of asking first whether fits to diabatic dimeson couplings alone are sufficient to obtain this spectrum (and as seen in Sec.~\ref{sec:spindependentmatrices}, they are not).  Then, using the best of such fits as a starting point, we incorporate the parameters of the adiabatic model to achieve a perfect fit to the known masses, and predict masses for the remainder of the $\Sigma^+_g(1S)$ multiplet.  In this approach, the $\Sigma^+_g(1S)$ spectrum originates from the discrete spectrum of the confining \(\de\bde\) channel, whose eigenstates are subsequently dressed through coupling to the available dimeson channels.  States lying below the relevant dimeson thresholds remain bound,\footnote{The actual physical states remain unstable: Decays through channels like charmonium plus light hadrons remain.} while those embedded in an open-channel continuum acquire a finite decay width and appear as resonances, with their masses shifted slightly from the corresponding uncoupled \(\de\bde\) eigenvalues~\cite{Lebed:2024rsi}.

% In this approach, the $\Sigma^+_g(1S)$ spectrum is treated as being largely, but not completely, determined by dimeson threshold couplings, with $\de\bde$ couplings in distinct spin/isospin channels introduced only as necessary to fit the known states.

In our fits, the diabatic parameters \(\{\theta, \, \Delta, \, \rho, \, m_{\de}\}\) are first fixed to the representative best set of the pure spin-dependent diabatic search listed in Eq.~\eqref{eq:bestdiabaticparameters}.  Holding these parameters fixed, we then calibrate the potential by obtaining the minimum correction parameters $C$ for the \(\de\bde\) channel potential needed to recover the known masses [Eq.~\eqref{eq:PDGmasses}] of the experimentally identified states \(\chi_{c1}(3872)\), \(T_{c\bar{c}1}(3900)\), and \(T_{c\bar{c}1}(4020)\).

The resonance structure associated with each state is obtained by identifying its experimentally observed open-charm decay modes~\cite{ParticleDataGroup:2026mpi}, constructing the scattering amplitudes for the corresponding two-body dimeson channel in the model, and examining features in this amplitude.  For \(\chi_{c1}(3872)\), the observed \(D^{0}\bar{D}^{0}\pi^{0}\) final state proceeds through the near-threshold \(D^{0}\bar{D}^{\ast0}+\bar{D}^{0}D^{\ast 0}\) channel, represented here by \(D\bar{D}^{\ast}\).  The corresponding channels for \(T_{c\bar{c}1}(3900)\) and \(T_{c\bar{c}1}(4020)\) are \(D\bar{D}^{\ast}\) and \(D^{\ast}\bar{D}^{\ast}\), respectively. The correction \(C \to C_{n}\) to the \(\de\bde\) channel for the $n^{\rm th}$ state is then varied until the resonant peak in this channel coincides with the experimental mass of the corresponding state.  Hence, this calibration strategy does not rely solely upon selecting the resonance structure nearest to each experimental mass, but additionally uses the observed decay channels to associate the known state with the appropriate cross-section structure.

We use an initial coarse variation of \(C_{n}\) to locate the relevant region for each state, followed by a finer scan in 0.1~MeV increments about the identified value.  Resonance locations are determined using the maxima of the corresponding elastic cross section,\footnote{In principle, one could instead extract and use the pole of each (complex) amplitude.} yielding values for the corrections $C_n$, $n$ labeled by the mass eigenvalue:

\begin{align}\label{eq:calibrationcorrections}
    C_{3872}&= \; -98.3~\mathrm{MeV},\notag\\ % -98.25
    C_{3900}&= \; -85.8~\mathrm{MeV},\notag\\
    C_{4020}&= \; +50.5~\mathrm{MeV}.
\end{align}
These results provide clear indications of the inadequacy of the pure diabatic calculation, and parametrize the magnitude of the deficiency in each case.

As noted above, the corrections listed in Eq.~\eqref{eq:calibrationcorrections} are not three independent parameters of the final model; rather, they provide constraints on the adiabatic Hamiltonian that determine the adiabatic parameters \{\(\Delta M_{0}\), \(\kappa_{qQ}\), \(V_{0}\)\}.  Using Eq.~\eqref{eq:adiabaticmasssplittingtilde}, one can express the corrections corresponding to the three known states in terms of these adiabatic parameters, after applying the appropriate basis rotation with \(\theta \to \theta_{1^{+-}}^{I=1}\) to the mixed \(I=1\), \(J^{PC}=1^{+-}\) states.  The resulting system of equations is:
\begin{equation}
    \begin{pmatrix}
        C_{3872}\\
        C_{3900}\\
        C_{4020}
    \end{pmatrix} =
    \begin{pmatrix}
        1 & -1 & -3\\
        1 & -\cos\left(2\theta\right) & -1-2\sin\left(2\theta\right)\\
        1 & +\cos\left(2\theta\right) & -1+2\sin\left(2\theta\right)
    \end{pmatrix}
    \begin{pmatrix}
        \Delta M_{0}\\
        \kappa_{qQ}\\
        V_{0}
    \end{pmatrix} .
\end{equation}

Solving these equations yields the solutions for the adiabatic parameters:
\begin{align}\label{eq:correctionadiabaticparameters}
    \Delta M_{0}&= 16.09\ \mathrm{MeV}, \notag\\
    \kappa_{qQ}&=13.12\ \mathrm{MeV}, \notag\\
    V_{0}&=33.74\ \mathrm{MeV}.
\end{align}
% Was: 8.727, 12.896, 33.777
% New: 16.089240, 13.121520, 33.739240

Although \(\Delta M_{0}\), \(\kappa_{qQ}\), and \(V_{0}\) are now determined, the adiabatic correction matrices of Eq.~\eqref{eq:adiabaticmasssplittingtilde} are not yet fixed for the entire multiplet.  The mixing angle \(\theta_{1^{+-}}^{I=1}\) is already known from the calibration to \(T_{c\bar{c}1}(3900)\) and \(T_{c\bar{c}1}(4020)\) [Eq.~\eqref{eq:bestdiabaticparameters}]:
\begin{align}\label{eq:diabaticfittedangle}
    \theta_{1^{+-}}^{I=1}=0.624\ \mathrm{rad}.
\end{align}
However, no corresponding experimental information is available to independently determine the mixing angles in the remaining sectors.  These angles are unambiguously determined in the adiabatic model using the Hamiltonian Eq.~\eqref{eq:adiabaticspinhamiltonian}, since it produces full mass eigenstates from the mixed sectors for any given set \{$\Delta M_0, \kappa_{qQ}, V_0$\}.  Different threshold corrections for different states in the diabatic model may shift the mixing angles slightly, but since (as found below) the adiabatic parameters appear to dominate the fine structure of the mass spectrum, it is reasonable to assume that the same holds for the mixing angles.

The correction matrices \(\Delta\tilde{\vv{M}}\) of Eq.~\eqref{eq:adiabaticmasssplittingtilde} are expressed in terms of the heavy/light quark-spin basis of Eq.~\eqref{eq:tildebasis}.  The \(0^{++}\) basis is ordered as \(\left(\tilde{X}_{0},\tilde{X}_{0}'\right)\), and the \(1^{+-}\) basis is ordered as \(\left(\tilde{Z},\tilde{Z}'\right)\), where the first and second state of each pair [as in Eq.~\eqref{eq:adiabaticmasssplittingtilde}] has heavy-quark spin eigenvalue \(s_{c\bar{c}}=0,1\), respectively.

The diabatic potentials are instead constructed in terms of diquark-spin eigenstates \((X_{0},X_{0}')\) and \((Z,Z')\) from Eq.~\eqref{eq:defaultbasis}. The constant matrices \(\vv{U}_{0^{++}}\) and \(\vv{U}_{1^{+-}}\) from Eq.~\eqref{eq:xi0rotatedbasis} must therefore first be used to express the adiabatic Hamiltonian Eq.~\eqref{eq:adiabaticspinhamiltonian} in terms of the diquark-spin basis:
\begin{equation}\label{eq:adiabaticH0}
    \vv{H}_{0}=\vv{U}^{\dagger}_{J^{PC}}~\Delta\tilde{\vv{M}}~\vv{U}_{J^{PC}}=\begin{pmatrix}
        a & b\\
        b & d
    \end{pmatrix}.
\end{equation}
Since the remaining diabatic mixing angles (as noted above) are not experimentally constrained in the full analysis, we identify the corresponding rotated states with the mass eigenstates predicted by the adiabatic Hamiltonian Eq.~\eqref{eq:adiabaticspinhamiltonian}.  The transformation of \(\vv{H}_{0}\) into the corresponding \((\zeta,\zeta')\) [Eq.~\eqref{eq:zetarotatedbasis}] or (\(\xi_{0},\xi_{0}')\) [Eq.~\eqref{eq:xi0rotatedbasis}] basis reads:
\begin{equation}
    \vv{H}=\vv{R}\left(\theta_{J^{PC}}^{I}\right)\vv{H}_{0}\,\vv{R}^{\dagger}\!\left(\theta_{J^{PC}}^{I}\right).
\end{equation}
The value of \(\theta_{J^{PC}}^{I}\) is chosen to diagonalize \(\vv{H}\).  For the real symmetric matrix of Eq.~\eqref{eq:adiabaticH0}, one has:
\begin{equation}\label{eq:tangentadiabatixmixingangle}
    \tan\left(2\theta_{J^{PC}}^{I}\right)=-\frac{2b}{a-d}.
\end{equation}
The branch in Eq.~\eqref{eq:tangentadiabatixmixingangle} is chosen to restrict \(\theta_{J^{PC}}^{I}\in[0,\pi/2]\).  For the three unconstrained mixed sectors, one obtains the mixing angles (in rad):
\begin{align}
    \theta_{0^{++}}^{I=0} &= 0.471 \ = 0.300 \, \frac{\pi}{2},\notag\\ %  0.471075579647
    \theta_{0^{++}}^{I=1} &= 0.722 \ = 0.459 \, \frac{\pi}{2},\notag\\ % 0.721608537708
    \theta_{1^{+-}}^{I=0} &= 0.818 \ = 0.521 \, \frac{\pi}{2}.  % 0.817762035543
\end{align}
The resulting states may also be characterized by their heavy-quark spin content.  Combining the two basis transformations gives
\begin{equation}\label{eq:heavequarkspincontentmatrix}
    \vv{W}_{J^{PC}}^{I} \equiv \vv{R} \! \left(\theta^{I}_{J^{PC}}\right)\vv{U}^{\dagger}_{J^{PC}}.
\end{equation}
The rows of \(\vv{W}_{J^{PC}}^{I}\) correspond to the unprimed and primed states, while the columns correspond to the \(s_{c\bar{c}}=0,1\) heavy/light-spin basis states, respectively.  The corresponding state-content probabilities are therefore given by $|[\vv{W}^{I}_{J^{PC}}]^{\vphantom\dagger}_{ij}|^{2}$.

\begin{figure}[t]
    \centering
    \includegraphics[width=\columnwidth]{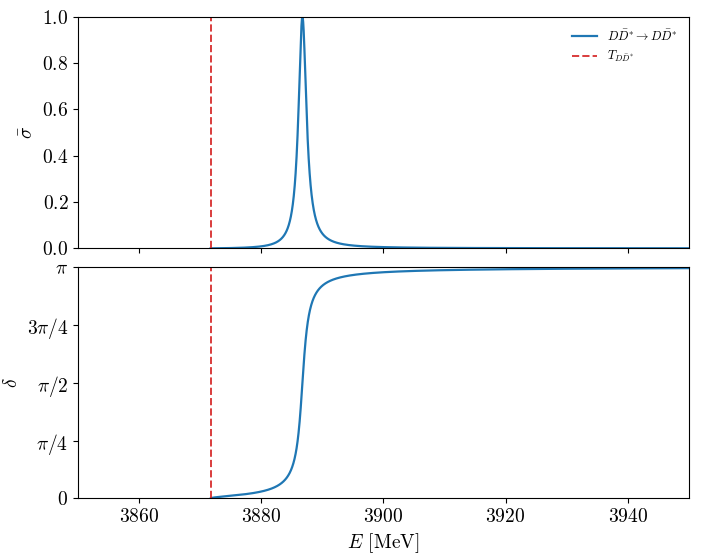}
    \caption{Normalized cross section \(\bar{\sigma}\) [Eq.~\eqref{eq:normcrosssection}] and \(D\bar{D}^{\ast}\) phase shift $\de$ for \(T_{c\bar{c}1}(3900)\).  The $D\bar D^*$ threshold is depicted by a dashed vertical line.}
    \label{fig:Tccbar1_3900}
\end{figure}

With the adiabatic parameters and mixing angles determined, the corrections for all remaining members of the \(\Sigma_{g}^{+}(1S)\) multiplet are now fixed.  The corresponding coupled-channel Schr\"{o}dinger equations may then be solved numerically to search for both bound states below the relevant dimeson thresholds\footnote{As noted above, such bound states are only stable against decays into open-heavy-flavor meson pairs.} and resonant structures in the normalized scattering cross sections \(\bar{\sigma}\) [Eq.~\eqref{eq:normcrosssection}] above threshold.  When a bound state is present, its energy is obtained directly from the numerical bound-state solution; otherwise, resonant structures are identified from the calculated cross sections, and are fit to Breit-Wigner distributions to determine their locations. A representative scattering solution is shown in Fig.~\ref{fig:Tccbar1_3900}, in which the resonant enhancement of the normalized cross section is accompanied by the characteristic rapid increase of the \(D\bar{D}^{\ast}\) phase shift $\delta$ through $\pi/2$. The remaining above-threshold members of the multiplet exhibit qualitatively similar resonant behavior. 

The resulting heavy-quark-spin state compositions and resonance locations are summarized in Table~\ref{tab:corrected_1S_probabilities_resonances}\@.  The mixed states are found to be strongly dominated by a single $s_{c\bar c}$ spin configuration, while the unmixed \(X_{1}\) and \(X_{2}\) states by Eq.~\eqref{eq:tildebasis} have only \(s_{c\bar{c}}=1\) components.

\begin{table*}[t]
    \centering
    \begin{tabular*}{\textwidth}{@{\extracolsep{\fill}}c c c c c c c c}
        \hline\hline
        $I$ & $J^{PC}$ & State
        & \multicolumn{2}{c}{{Heavy-quark spin state content}}
        & \multicolumn{3}{c}{Mass (MeV)} \\
        \cline{4-5}\cline{6-8}
        & & 
        & \phantom{---}$s_{c\bar {c}}=0$\phantom{---}
        & \phantom{---}$s_{c\bar {c}}=1$\phantom{---}
        & $D\bar D$
        & $D\bar D^{*}$
        & $D^{*}\bar D^{*}$ \\
        \hline

        \multirow{6}{*}{$0$}

        & \multirow{2}{*}{$0^{++}$}
        & $\xi_0$
        & 0.0028 & 0.9972
        & 3861.5 & $\ldots$ & $\ldots$ \\

        &
        & $\xi_0'$
        & 0.9972 & 0.0028
        & 4296.9 & $\ldots$ & 4296.9 \\

        & \multirow{2}{*}{$1^{+-}$}
        & $\zeta$
        & 0.0010 & 0.9990
        & $\ldots$ & 4296.2 & 4296.2 \\

        &
        & $\zeta'$
        & 0.9990 & 0.0010
        & $\ldots$ & 3886.6 & $\ldots$ \\

        & $1^{++}$
        & $X_1$
        & 0 & 1
        & $\ldots$ & \textbf{3871.7} & $\ldots$ \\

        & $2^{++}$
        & $X_2$
        & 0 & 1
        & $\ldots$ & $\ldots$ & \(3901.4^{\dagger}\) \\

        \hline

        \multirow{6}{*}{$1$}

        & \multirow{2}{*}{$0^{++}$}
        & $\xi_0$
        & 0.0387 & 0.9613
        & 4003.4 & $\ldots$ & $\ldots$ \\

        &
        & $\xi_0'$
        & 0.9613 & 0.0387
        & 3883.9 & $\ldots$ & $\ldots$ \\

        & \multirow{2}{*}{$1^{+-}$}
        & $\zeta$
        & 0.0259 & 0.9741
        & $\ldots$ & \textbf{3886.7} & $\ldots$ \\

        &
        & $\zeta'$
        & 0.9741 & 0.0259
        & $\ldots$ & 4023.8 & \textbf{4024.1} \\

        & $1^{++}$
        & $X_1$
        & 0 & 1
        & $\ldots$ & 4011.3 & $\ldots$ \\

        & $2^{++}$
        & $X_2$
        & 0 & 1
        & $\ldots$ & $\ldots$ & 4034.1 \\

        \hline\hline
    \end{tabular*}
    \caption{Heavy-quark-spin state content and scattering-amplitude peak locations (in MeV) of the lowest resonance structures for the diabatic-corrected $\Sigma_g^+(1S)$ multiplet.  Heavy-quark-spin state contents are given for each mass eigenstate, while resonance peak locations are reported for each open dimeson channel in which a structure is observed.  When a bound state occurs below threshold, its bound-state energy is reported instead, and is denoted by \(^{\dagger}\).  Bold entries denote input states of the model.}
    \label{tab:corrected_1S_probabilities_resonances}
\end{table*}

The distinction between the bound and resonant solutions may be understood by comparison with the corresponding uncoupled \(\de\bde\) spectrum.  In the absence of mixing with dimeson states, the confining Cornell-like potential Eq.~\eqref{eq:diquarkpotential} supports a discrete spectrum of states.  Coupling to the dimeson channels dresses these states, shifting their locations slightly and, when they lie above an open threshold, allows them to acquire a finite partial width in that channel.  The eigenvalues of the uncoupled \(\de\bde\) system therefore provide a useful indication of where the corresponding states appear in the coupled problem.  In particular, the \(I=0\), \(J^{PC}=2^{++}\) state lies near \(3904\) MeV in the uncoupled \(\de\bde\) spectrum, below its lowest $S$-wave open-heavy-flavor dimeson threshold \(D^{\ast}\bar{D}^{\ast}\) at about 4014 MeV\@.  It consequently appears as a bound-state solution within the present coupled-channel basis.  This does not imply that the physical state is stable, since decay channels not included in the model ({\it e.g.}, $J/\psi \, \omega$, or $D\bar D$ in a $D$-wave) may remain kinematically accessible, and generate a finite width.  The corresponding levels for the remaining multiplet members lie above an available $S$-wave dimeson threshold, and instead appear as resonant structures in the coupled-channel calculation.

The state reported for each member of the multiplet is taken to be the lowest-energy bound state when one is present; otherwise the lowest-energy resonance is used.  For states coupling to multiple  $S$-wave dimeson channels, approximately coincident structures appearing in different channels are regarded as manifestations of the same coupled-channel resonance.  Additional resonances observed at higher energies are not included here, and presumably indicate higher excitations of the associated state.  The resulting spectrum preserves the mass ordering obtained from the pure adiabatic Hamiltonian: No inversion of the adiabatic ordering is observed for any of the mixed ($0^{++}$, $1^{+-}$) pairs.

The \(\chi_{c1}(3872)\), \(T_{c\bar{c}1}(3900)\), and \(T_{c\bar{c}1}(4020)\) structures in Table~\ref{tab:corrected_1S_probabilities_resonances} reproduce their experimental masses by construction, and therefore do not constitute independent predictions of the model.  The remaining entries, however, follow from the same calibrated parameter set, without the introduction of additional state-dependent corrections.  An interesting point of comparison arises from the spin-dependent parameters obtained in this calibration.  The values of \(\kappa_{qQ}\) and \(V_{0}\) in Eq.~\eqref{eq:correctionadiabaticparameters} are numerically quite comparable to
the best-fit values using the same adiabatic Hamiltonian [but setting $V_8 =0$ in Eq.~\eqref{eq:adiabaticspinhamiltonian}] that are given in Eq.~(18) of Ref.~\cite{Giron:2020qpb}:
\begin{align}\label{eq:pureadiabaticparameters}
    M_{0}&=3988.7\ \mathrm{MeV}, \notag\\ % I reference these values together for comparison later
    \kappa_{qQ}&=17.89\ \mathrm{MeV}, \notag\\
    V_{0}&=33.04\ \mathrm{MeV}.
\end{align}
The remarkable numerical stability of $\kappa_{qQ}$ and $V_{0}$ in the two fits suggests that, rather than the pure diabatic correction serving as the dominant source of the fine structure of $\Sigma^+_g(1S)$, it is in fact the (iso)spin-dependent interactions of the $\de\bde$ pair that determines most of the mass hierarchy.  In particular, \(V_{0}\) differs by less than 1 MeV between the two determinations, while \(\kappa_{qQ}\) remains of comparable magnitude (about 40\% larger).

The common terms \(M_{0}\) and \(\Delta M_{0}\), however, have radically different interpretations: \(M_{0}\) sets the common multiplet mass in the pure adiabatic model, whereas \(\Delta M_{0}\) enters the present construction as a common shift to the \(\delta\bar{\delta}\) channel potential compared to the $D^{(*)}$ spin-averaged mass $m_0$ in Eq.~\eqref{eq:DDspinavemass}.  Since \(\kappa_{qQ}\) and \(V_{0}\) produce fine-structure mass splittings from the same underlying Hamiltonian [Eq.~\eqref{eq:adiabaticspinhamiltonian}], the numerical similarity of Eqs.~\eqref{eq:correctionadiabaticparameters} and \eqref{eq:pureadiabaticparameters} motivates a comparison of the fine-structure patterns produced by the two calculations.  The complete \(c\bar{c}q\bar{q}'\) \(\Sigma^+_g(1S)\) spectrum using the parameters of Eq.~\eqref{eq:pureadiabaticparameters} is not tabulated explicitly in Ref.~\cite{Giron:2020qpb}, although very similar sets of predictions, obtained under slightly different assumptions, are tabulated in Refs.~\cite{Giron:2019bcs} and \cite{Giron:2021sla}.  The remaining masses nevertheless follow directly from the adiabatic mass expressions given in Eq.~\eqref{eq:adiabaticmasssplittingcorr} and Ref.~\cite{Giron:2020qpb}.  Evaluating those expressions yields the spectrum shown in Table~\ref{tab:pureadiabatic1S}.

\begin{table}[t]
    \centering
    \begin{tabular}{c c c c}
        \hline\hline
        \hspace{1em} $I$ \hspace{1em} & $J^{PC}$ & Eigenstate & Mass (MeV) \\
        \hline

        \multirow{6}{*}{$0$ \ }
        & \multirow{2}{*}{$0^{++}$}
        & $M_{-}$ & 3851.6 \\
        &
        & $M_{+}$ & 4288.3 \\

        & \multirow{2}{*}{$1^{+-}$}
        & $M_{-}$ & 3888.8 \\
        &
        & $M_{+}$ & 4286.9 \\

        & $1^{++}$
        & $X_1$ & \textbf{3871.7} \\

        & $2^{++}$
        & $X_2$ & 3907.5 \\

        \hline

        \multirow{6}{*}{$1$ \ }
        & \multirow{2}{*}{$0^{++}$}
        & $M_{-}$ & 3880.5 \\
        &
        & $M_{+}$ & 3995.1 \\

        & \multirow{2}{*}{$1^{+-}$}
        & $M_{-}$ & \textbf{3887.2} \\
        &
        & $M_{+}$ & \textbf{4024.1} \\

        & $1^{++}$
        & $X_1$ & 4003.8 \\

        & $2^{++}$
        & $X_2$ & 4039.6 \\

        \hline\hline
    \end{tabular}
    \caption{Mass spectrum of the \(\Sigma_g^+(1S)\) \(c\bar{c}q\bar{q}'\) multiplet obtained from the pure adiabatic model, using the parameters of Ref.~\cite{Giron:2020qpb}.  For the mixed \(0^{++}\) and \(1^{+-}\) sectors, \(M_{-}\) and \(M_{+}\) denote the lower and higher mass eigenstates, respectively.  Bold entries again denote the states used in calibrating the model parameters; note that the $I=1$, $J^{PC}=1^{+-}$ experimental input shifted slightly since the publication of Ref.~\cite{Giron:2020qpb}.}
    \label{tab:pureadiabatic1S}
\end{table}

A state-by-state comparison of the two spectra (in Tables~\ref{tab:corrected_1S_probabilities_resonances} and \ref{tab:pureadiabatic1S}) requires some care, since the primed and unprimed mixed-state labels do not correspond uniformly to the lower ($M_-$) and upper ($M_+$) eigenvalues of the pure adiabatic Hamiltonian Eq.~\eqref{eq:adiabaticspinhamiltonian}.  The correspondence may instead be established from the heavy-quark spin compositions of the states, expressed in the common \(s_{c\bar{c}}=0,1\) basis.  This identification gives:

\begin{align}
    I=0,\quad J^{PC} = 0^{++}:&\qquad (M_-,M_+)=(\xi_0,\xi_0'),\notag\\
    I=0,\quad J^{PC} = 1^{+-}:&\qquad (M_-,M_+)=(\zeta',\zeta),\notag\\
    I=1,\quad J^{PC} = 0^{++}:&\qquad (M_-,M_+)=(\xi_0',\xi_0),\notag\\
    I=1,\quad J^{PC} = 1^{+-}:&\qquad (M_-,M_+)=(\zeta,\zeta').
\end{align}
For the first three of these mixed sectors, the mixing angles \(\theta_{J^{PC}}^{I}\) used in the diabatic calculation are those obtained by diagonalizing the corresponding adiabatic mass matrices, as obtained from Eq.~\eqref{eq:tangentadiabatixmixingangle}. Consequently, these states have heavy-quark-spin compositions identical to those obtained for the corresponding \(M_{\pm}\) eigenstates in the pure adiabatic calculation~\cite{Giron:2020qpb}. The \(I=1\), \(J^{PC}=1^{+-}\) sector differs in that its mixing angle is determined independently, using the diabatic fit with the parameter values of Eq.~\eqref{eq:bestdiabaticparameters}, and therefore does not exactly diagonalize the corresponding adiabatic mass matrix. The diabatic fitted angle 0.624~rad [Eq.~\eqref{eq:diabaticfittedangle}] nevertheless remains close to the adiabatic diagonalizing value of \(0.698\) rad. Hence, the resulting \(\zeta\) and \(\zeta'\) states may still be identified with the adiabatic \(M_{-}\) and \(M_{+}\) eigenstates, respectively, to a high degree of accuracy. 

Comparison with Table~\ref{tab:corrected_1S_probabilities_resonances} shows that the numerical similarity of the (iso)spin-dependent parameters is accompanied by a striking similarity in the resulting spectra.  For all members of the multiplet, the lowest mass eigenvalues obtained from the corrected diabatic calculation lie no more than 10 MeV from the corresponding pure adiabatic mass eigenvalues.

In both constructions, the fine structure originates primarily from the same spin-dependent Hamiltonian Eq.~\eqref{eq:adiabaticspinhamiltonian}, and the comparable values of \(\kappa_{qQ}\) and \(V_{0}\) therefore generate a similar underlying pattern of mass splittings.  Despite this common underlying fine structure, the determination of the physical spectrum differs fundamentally between the two approaches: The pure adiabatic model obtains a discrete spectrum directly, whereas in the diabatic construction, the corresponding corrections modify the \(\delta\bar{\delta}\) channel potential before the full coupled-channel scattering is solved.  The adiabatic model may therefore be viewed as establishing much of the underlying fine structure, while the full coupled-channel calculation determines how these configurations appear after mixing with the open dimeson thresholds.

\section{Discussion} \label{sec:Disc}

Beyond the states \(\chi_{c1}(3872)\), \(T_{c\bar{c}1}(3900)\), \(T_{c\bar{c}1}(4020)\) used in the calibration, the results for the remaining resonance structures constitute predictions of the corrected diabatic model, and may therefore be compared with other observed structures in the hidden-charm spectrum.  No unique assignments have been established for the remaining members of the \(\Sigma_g^+(1S)\) multiplet, although several experimentally observed structures have nevertheless been proposed as possible candidates in previous tetraquark studies.  However, in many cases their quantum numbers or even existence remains uncertain.  The following comparisons are thus intended only to suggest potential members of the $\Sigma^+_g(1S)$ multiplet.

\(\chi_{c0}(3860)\) has been considered as an \( I^G = 0^{+}, \, J^{PC} = 0^{++}\) tetraquark candidate in Refs.~\cite{Giron:2020qpb,Shi:2021jyr}, and its mass ${3862}^{+50}_{-35}$~MeV~\cite{ParticleDataGroup:2026mpi} lies remarkably close to the corresponding resonance at approximately 3861.5 MeV in Table~\ref{tab:corrected_1S_probabilities_resonances}\@.  Its interpretation however, remains unsettled, as it could also be the conventional charmonium $\chi_{c0}(2P)$ or a mixture with the tetraquark; hence, mass agreement alone is insufficient to establish its membership in $\Sigma^+_g(1S)$.  A similar situation occurs for the isovector \(2^{++}\) state at 4033.8~MeV\@.  The unconfirmed \(Z_{c}^+(4055)\)~\cite{Belle:2014wyt} was noted~\cite{Giron:2019cfc} as a possible \(X_{2}^{I=1}\) candidate, while Ref.~\cite{Shi:2021jyr} considered \(Z_c^+(4050)\) to have the same quantum numbers.  Both lie within approximately 15~MeV of the present result, although these states again remain insufficiently established for direct identification.

More recently, the observation of \(h_{c}(4300)\) with \(J^{PC}=1^{+-}\)~\cite{LHCb:2024vfz} provides a possible candidate for the higher $I=0$, \(J^{PC}=1^{+-}\) state, obtained here near 4296.2 MeV, although of course a conventional charmonium assignment [specifically, $h_c(3P)$] is also possible.
%Additionally, Ref.~\cite{LHCb:2024vfz} also observes a state labeled as $\chi_{c1}(4010)$, but note that they do not measure isospin, meaning that this state would agree well with our predicted $I_3 = 0$ $X_1^{I=1}$ state.  Unfortunately, since $C=+$ and $G=-$ for this state, it cannot decay into $D^* \bar D + {\rm h.c.}$}
In contrast to several of the older candidates, the measured $J^{PC}$ assignment of $h_c(4300)$ agrees with that of the predicted state, making it a particularly interesting tetraquark candidate.

The remaining possibilities are considerably more ambiguous.  \(X(3940)\), for example, has received several tetraquark assignments, including both \(I=1\), \(J^{PC}=0^{++}\) and \(I=1\), \(J^{PC}=1^{++}\)~\cite{Giron:2019cfc,Shi:2021jyr}.  Consequently, its proximity to more than one member of the calculated multiplet cannot presently distinguish among these possibilities, and moreover, both predictions appear somewhat too high to accommodate the \(X(3940)\) mass.  Similar qualifications apply to the proposed assignments of \(X(4160)\), \(Z_{c}(4100)\), \(Z_{c}(4200)\), and \(Z_c(4250)\) to $\Sigma^+_g(1S)$, since either the relevant quantum numbers or existence of the state remain insufficiently determined~\cite{Giron:2019cfc,Yu_2024}.

The \(I=0\), \(J^{PC}=2^{++}\) state is of particular interest in connection with \(\chi_{c2}(3930)\), which has previously been considered as a possible isoscalar \(2^{++}\) member of the multiplet.  In the present calculation, the lowest state in this sector occurs as a bound-state solution near \(3.90\) GeV, rather than as a resonance, found much higher near \(4.48\)~GeV\@. The latter can instead be associated with a higher ($2S$) excitation of the \(\de\bde\) spectrum.  Although \(\chi_{c2}(3930)\) is generally interpreted predominantly as conventional charmonium \(\chi_{c2}(2P)\), its quantum numbers permit mixing with the \(I=0\), \(J^{PC}= 2^{++}\) tetraquark configuration considered here.  Moreover, the bound-state character of the calculated state only applies with respect to the dimeson channels included in the present model.  In particular, the \(D\bar{D}\) channel, through which the \(\chi_{c2}(3930)\) is observed and that can contribute to this state in a \(D\) wave, is absent from the present channel basis (although it certainly can be included~\cite{Lebed:2023kbm,Lebed:2025eqm}).  Incorporating this channel, together with possible mixing with conventional charmonium, could therefore generate a finite width and modify the location and composition of the state.  A more complete coupled-channel treatment would be required to determine whether the state obtained here is related to the physical \(\chi_{c2}(3930)\), and thus no direct identification is made at present.

The available experimental structures provide several suggestive comparisons, but do not presently permit unique assignments for the remaining members of the multiplet.  The closest correspondences are best regarded as possible targets for further experimental study rather than as concrete identifications.  Improved determinations of \(J^{PC}\), isospin, and decay modes would be particularly valuable for distinguishing among the competing assignments.  Furthermore, the present calculation is constrained by only a limited set of experimental inputs, resulting in the parameter degeneracy discussed previously.  The resonance locations in Table~\ref{tab:corrected_1S_probabilities_resonances} should therefore be understood as predictions conditioned upon the particular calibrated parameter set obtained here; additional experimental constraints could select a somewhat different region of parameter space and consequently modify the predicted spectrum, although the strong similarity of the adiabatic and diabatic spectra suggests that many of the predictions are robust.

Lastly, the number of ambiguous candidates already appearing in the expected mass range also suggests that the eventual discovery of a full spectrum of 12 states in the $\Sigma^+_g(1S)$ multiplet is not an unreasonable expectation.

%\note{Particle names here were used from their respective papers.  Some particles have changed a lot for some reason; for example I don't actually see a listing for the \(Z_{c}(4100)\), I think at some point it changed to \(X(4100)^{\pm}\) and then that changed again to \(T_{c\bar{c}}(4100)^{+}\).  Should I go hunting for the modern names for these particles or keep them as-is? I am unsure if this kind of thing is common knowledge for people in the business.}

\section{Conclusions}\label{sec:Conclusion}

In this paper, the HQSS-constrained diabatic dynamical diquark model was used to describe the hidden-charm ground-state \(\Sigma_{g}^{+}(1S)\) multiplet.  A direct scan of the original diabatic parameter space [{\it i.e.}, ignoring corrections to the diquark-antidiquark ($\de\bde$) potential] was found insufficient for simultaneously reproducing the measured \(\chi_{c1}(3872)\), \(T_{c\bar{c}1}(3900)\), and \(T_{c\bar{c}1}(4020)\) resonance masses.  This result mandates the incorporation of the fine structure predicted by the adiabatic dynamical diquark model into the compact \(\delta\bar{\delta}\) channel of the diabatic potential.

These three experimentally identified states were then used to calibrate the model by determining the {\em minimal} \(\delta\bar{\delta}\) spin- and isospin-dependent corrections necessary to reproduce their spectrum, from which the adiabatic parameters \(\Delta M_{0}\), \(\kappa_{qQ}\), and \(V_{0}\) were obtained.  The remaining experimentally unconstrained mixing angles were then supplied by the same mass-eigenstate mixing predicted by the corresponding pure adiabatic Hamiltonian.  This procedure fixes the mass eigenvalues and mixing angles in the remaining members of the multiplet without introducing additional fit parameters.

Solving the resulting coupled-channel Schr\"{o}dinger equations produces resonance structures for the full \(\Sigma_{g}^{+}(1S)\) multiplet (with one state remaining as bound under $S$-wave heavy-open-flavor decays).  The three calibration states are reproduced by construction, while the remaining results constitute predictions of the model.  The mixed \(0^{++}\) and \(1^{+-}\) states were each found to be strongly dominated by a single heavy-quark spin configuration, providing an approximate heavy-spin-eigenstate characterization of the resulting states.

One particularly remarkable result arises from comparison with the pure adiabatic model.  The fitted values of \(\kappa_{qQ}\) and \(V_{0}\) are similar to those obtained in the purely adiabatic analysis of Ref.~\cite{Giron:2020qpb}, and this similarity is reflected in the resulting spectra: All of the calculated resonance locations lie within approximately \(10~\mathrm{MeV}\) of the corresponding pure-adiabatic mass eigenvalues.  This surprising finding suggests that much of the $\Sigma^+_g(1S)$ mass spectrum is established by the adiabatic (iso)spin interactions, and survives the introduction of explicit couplings to dimeson channels. 

Several of the predicted structures also lie near experimentally reported charmoniumlike states that have previously been discussed as tetraquark candidates.   Of particular interest are the lighter $I=0$, \(J^{PC}=0^{++}\) resonance near \(3862~\mathrm{MeV}\), which may be compared with \(\chi_{c0}(3860)\), and the heavier $I=0$, \(J^{PC}=1^{+-}\) structure near \(4297~\mathrm{MeV}\), for which the recently observed \(h_{c}(4300)\) invites comparison.  These identifications remain tentative, since the available experimental information is insufficient to establish unique assignments to the multiplet.

The present predictions remain limited by the small number of established states available for constraining the model.  The three calibration masses do not uniquely determine the underlying diabatic parameter set \(\{\theta, \, \Delta, \, \rho, \, m_{\de}\}\); alternate regions of parameter space may lead to different predictions for the unobserved members of the multiplet (although the remarkable stability of the adiabatic predictions argues against radical changes).  Additional measurements of masses, quantum numbers, decay modes, and resonance line shapes would provide crucial constraints on the remaining parameter space.

Nevertheless, the present construction demonstrates that the fine structure of the adiabatic dynamical diquark model can be incorporated into the spin-dependent diabatic framework, while maintaining the established description of the experimentally confirmed states.  These calculations extend this description to the complete hidden-charm \(\Sigma_{g}^{+}(1S)\) multiplet, and provide concrete resonance and bound-state locations along with heavy-quark spin characteristics for its remaining members.  Future experimental information on these states will therefore provide a direct test of both the predicted multiplet structure and the role of open-channel dynamics in the dynamical diquark model.

\begin{acknowledgments}
This work was supported by the National Science
Foundation (NSF) under Grant No.\ PHY-2405262.
\end{acknowledgments}

\bibliographystyle{apsrev4-2}
\bibliography{MasterRefs}

\end{document}